\documentclass[%
reprint,
superscriptaddress,
 amsmath,amssymb,
 aps,
 pra,
]{revtex4-1}

\usepackage{float}
\usepackage{graphicx}
\usepackage{dcolumn}
\usepackage{bm}

\usepackage[utf8]{inputenc} 

\usepackage{multirow}
\usepackage[warn]{mathtext}
\usepackage{amsfonts}
\usepackage{amsmath}
\usepackage{amssymb}
\usepackage{braket}
\usepackage{bbold}
\usepackage{textcomp} 
\usepackage{indentfirst} 
\usepackage{amsmath} 
\usepackage{graphicx}

\DeclareGraphicsExtensions{.pdf,.png,.jpg}

\usepackage{hyperref}

\usepackage{algpseudocode}

\usepackage[nottoc]{tocbibind}

\usepackage{xcolor}
\hypersetup{
    colorlinks,
    linkcolor={red!50!black},
    citecolor={blue!50!black},
    urlcolor={blue!80!black}
}



\begin{document}

    \title{Granular Aluminum kinetic inductance nonlinearity}

    \author{M.~Zhdanova}
    \email{zhdanova.mar@gmail.com}
    \affiliation{National University of Science and Technology MISIS, Moscow 119049, Russia}
 
    \author{I.~Pologov}
    \affiliation{National University of Science and Technology MISIS, Moscow 119049, Russia}

    \author{ G.~Svyatsky}
    \affiliation{National University of Science and Technology MISIS, Moscow 119049, Russia}
    
    \author{V.I.~Chichkov}
     \affiliation{National University of Science and Technology MISIS, Moscow 119049, Russia}

    \author{N.~Maleeva}
    \affiliation{National University of Science and Technology MISIS, Moscow 119049, Russia}

\date{\today}
 
	\begin{abstract}
	Granular Aluminum is a superconductor known for more than eighty years, which recently found its application in  qubits, microwave detectors and compact resonators, due to its high kinetic inductance, critical magnetic field and critical current. Here we report on the nonlinear dependence of granular Aluminum inductance on current, which hints towards parametric amplification of the microwave signal in granular Aluminum films. The phase shift of the microwave signal reached 4 radians at a frequency of 7~GHz, which makes it possible to estimate the nonlinearity of the system as $\Delta\phi/\phi = 1.4\ \%$ and the potential gain of the order of 17~dB.
    \end{abstract}
	
\maketitle

\section{Introduction} 
Superconductors are one of the most promising platforms for quantum processor realisation. Transmon-based processors have been used in first experiments on quantum supremacy demonstration \cite{Franson_2002, Arute_2019}, quantum error-correcting surface codes realizations \cite{Google_QAI_Acharya_2023, Krinner_2022} and solving quantum chemistry problems \cite{Google_QAI_Arute_2020, Kandala_2017}. Such processors require an amplification of a readout signal. Superconducting parametric amplifiers are the best in terms of signal-to-noise ratio, having an additional noise at the quantum limit level: 168~mK at 7~GHz \cite{Roy_2016}.

Traveling-wave parametric amplifiers (TWPA), used for a wide-band parametric amplification, employ either an array of Josephson junctions  \cite{Macklin_2015, Castellanos_Beltran_2007, Castellanos_Beltran_2008}, or a high kinetic inductance materials like NbTiN \cite{Zmuidzinas_2012, Malnou_2021, Chaudhuri_2017}, TiN \cite{Leduc_2010} or NbN \cite{Zhao_2023}. Josephson junctions fabrication consists of several complicated technological operations implementing a multi-layered structure of the junctions, which makes fabrication of a 1000 junction array \cite{Macklin_2015, Zorin_2017, Planat_2020, Ranadive_2022} a rather complex task. On the other hand TWPA based on high kinetic inductance films \cite{Vissers_2016, Shu_2021} reduce fabrication to one technological cycle and minimize production risks. Nevertheless a nonlinearity of such systems is concomitant with local heating and hot spots emerging, which leads to the structure's critical current decrease \cite{Skocpol_1974,Kurter_2011}.

Another material of choice for microwave cryogenic nonlinear devices realisations is granular Aluminum (grAl) \cite{Cohen_1968,Deutscher1973}. Due to its high kinetic inductance, up to 8~nH per square \cite{Gr_nhaupt_2018}, grAl is applicable in design of MKIDs \cite{Day_2003, Quaranta_2013, Szypryt_2015, Battistelli_2015, Cardani_2015, Cardani_2017, Valenti_2019, Levy_Bertrand_2021}, parametric amplifiers \cite{Zmuidzinas_2012, Vissers_2016}, fluxoniums \cite{Manucharyan_2009, Pop_2014, Lin_2018, Earnest_2018, Cohen_2017}, nonlinear resonators \cite{Rotzinger_2016}, filters \cite{Irwin_2010} and metamaterials~\cite{Kurter_2010}.  

Here we consider grAl as an alternative to traditional in TWPA realizations NbTiN \cite{Zmuidzinas_2012, Malnou_2021, Chaudhuri_2017, Vissers_2016}. The microstructure of grAl consists of pure aluminum grains separated by thin aluminum oxide barriers, forming a 3D-network of Josephson junctions, which results in high kinetic inductance \cite{Maleeva_2018}. Such a microstructure is obtained by evaporating of Aluminum in Oxygen atmosphere. An additional advantage of this material is a critical temperature on the order of 2–3~K (achievable by pumping He$^4$ and significantly higher then Aluminum $T_\mathrm{c}^\mathrm{Al} \approx$ 1.2~K), high critical fields \cite{Borisov_2020}, and handy stable fabrication technology.

\begin{figure}[htb]
\centering
\includegraphics[width=0.48\textwidth]{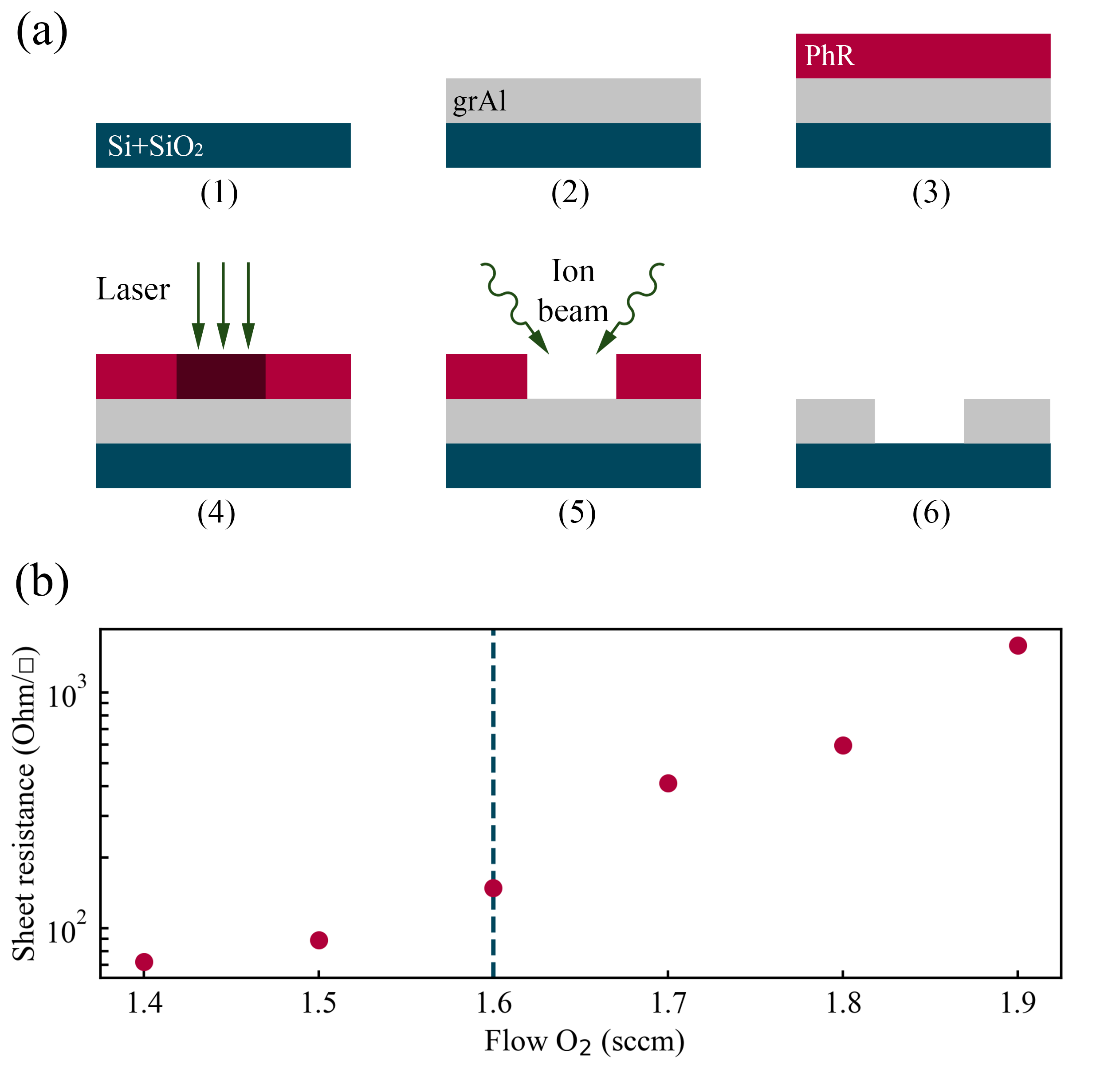}
\caption{\label{fig:1} (a) Stages of a sample fabrication: a grAl film is deposited by an electron beam evaporator (2) on an oxidized silicon substrate (1). After spin-coating a photoresist (3), a mask is formed by an optical lithography (4). Then by an ion gun etching (5) a final topology is created (6). 
(b) A calibration curve: an oxygen flow dependence of a grAl resistance per square. A vertical line indicates a sample with resistance per square of 200~$\Omega$, considered in this work.
}
\end{figure} 

\section{Fabrication} 
By adjusting an oxygen flow during Aluminum evaporation, one can control a resistance per square of a grAl film \cite{Pracht_2016}. Depending on the flow a resistivity varies between 10 and 10$^4~\mu\Omega\cdot\mathrm{cm}$ and critical temperature goes up to 3~K \cite{Pracht_2016}.

\begin{figure*}[tbh]
\centering
\includegraphics[width=0.98\textwidth]{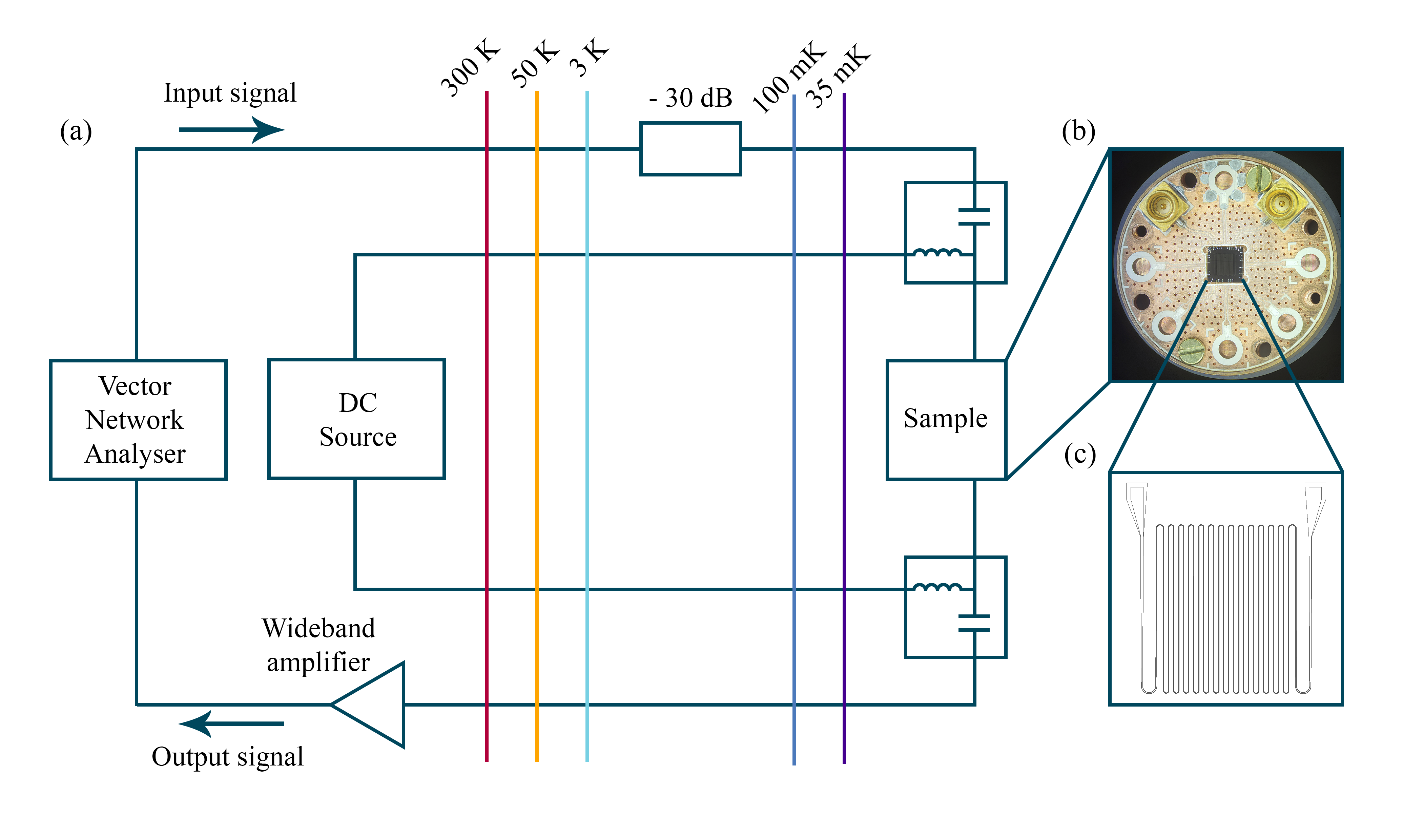}
\caption{\label{fig:2} An experimental setup scheme (a). In a dilution refrigerator a Copper sample holder (b) with a long coplanar line represented by a meander (c) is cooled down. We measure the transmission coefficient with a vector network analyzer (VNA) as a function of the dc current injected to the transmission line by bias-tees. An input signal is attenuated by 
30 dB and an output signal is amplified by a commercial wideband amplifier.}
\end{figure*}

The samples for this study were fabricated in an ISO-7 cleanroom using thin film planar technology methods.  Films were deposited by electron beam evaporator Plassys MEB 550 S. The samples topology was formed by an optical lithography using a photoresist MICROPOSIT S1805. After forming the protective mask, the substrate was placed back in Plassys MEB 550 S, where uncoated grAl areas were etched by an ion beam. This stages of sample fabrication are schematically shown in Fig.~\ref{fig:1}a. Finally, substrates were diced in 5x5 mm peaces and all organic residuals were washed off. 
Using test structures of etched bridges 8~$\mu$m wide and 200~$\mu$m long, the resistance per square for each oxygen flow value was measured by four-probe method resulting in a calibration curve shown in Fig.~\ref{fig:1}b. 
For this study we choose resistance of $R_\mathrm{n}^{\square}\ \approx$ 200~$\Omega$, corresponding to 1.6~sccm O$_2$ flow, for optimizition of the phase velocity and characteristic impedance of transmission line.

\section{Experimental study of granular Aluminum nonlinearity}

Considering thin grAl films we have a kinetic inductance $L_\mathrm{k}$  dominating over geometric $L_\mathrm{g}$. Moreover, for high resistances $R$ kinetic inductance can be several orders greater then geometric \cite{Mattis_1958}. A superconductor kinetic inductance can be found in the frame of the Bardeen–Cooper–Schrieffer (BCS) theory. In the limit of low frequencies ($hf \ll k_\mathrm{B} T_\mathrm{c}$) according to Mattis-Bardeen formula for dirty superconductors the complex conductivity can be written in terms of the ratio of the imaginary conductivity $\sigma_2$ to the
normal state conductivity $\sigma_\mathrm{n}$ as:
\begin{equation}\label{sigma2}
    \frac{\sigma_2}{\sigma_\mathrm{n}}=\frac{\pi \Delta}{h f} \tanh \frac{\Delta}{2 k_\mathrm{B} T},
\end{equation}
where $\Delta$ is the superconducting energy gap depending in general on temperature and bias current.
Introducing a kinetic inductance as $L_\mathrm{k} = 1/2\pi f \sigma_2$, one can write it down for a strip of length $l$ and width $w$
\begin{equation} \label{Lk}
L_\mathrm{k}  = N \frac{\hbar R_\mathrm{n}^{\square}}{\pi \Delta} \frac{1}{\tanh \frac{\Delta}{2 k_\mathrm{B} T}},
\end{equation}
where $N= l/w$ is a number of squares in a strip and $R_\mathrm{n}^{\square}$ is its sheet resistance in the normal state.
In the limit of low temperatures ($T\ll T_\mathrm{c}$), BCS formula for the energy gap  $\Delta=1.76 \pi k_\mathrm{B} T_\mathrm{c}$ can be used and the standard expression \cite{Rotzinger_2016} can be derived
\begin{equation}
\label{Lk at T=0}
L_\mathrm{k} = 0.18 N \frac{\hbar R_\mathrm{n}^{\square}}{k_\mathrm{B} T_\mathrm{c}}.
\end{equation}
Following a method given by \cite{Annunziata_2010} a kinetic inductance dependence on bias current at~$T~\ll~T_\mathrm{c}$ is considered as
\begin{equation}\label{Lk(I)}
 L_\mathrm{k} (I) = L_\mathrm{k}(0) e^{-\pi \zeta / 4},   
\end{equation}
where $L_\mathrm{k}(0)$ is given by Eq.~\eqref{Lk at T=0} and $\zeta = \frac{Dk^2}{2\Delta(I)}$ is a combination of the diffusion coefficient $D$, the gradient of the phase along the length of the superconducting strip $k$ and the enrgy gap $\Delta (I)$, corresponding to bias current $I$. For each bias current value $\zeta$ can be found by solving following equation 
\begin{equation}\label{I/Ic}
    \frac{I}{I_{\mathrm{dep}}}=1.897e^{-3\pi \zeta /8} \sqrt{\zeta}\left( \frac{\pi}{2} -\frac{2}{3}\zeta \right),
\end{equation}
where $I_\mathrm{dep}$ is the deparing current from BCS. Eqs.~\eqref{Lk(I)} and~\eqref{I/Ic} are obtained by solving Gorkov equation in the dirty superconductor limit \cite{Maki_1964}.
On the other hand a kinetic inductance in general nonlinearly depends on a bias current due to the Cooper pairs breaking and can be estimated from Pippard's equation as
\begin{equation}
\label{eq:2}
L_\mathrm{k}(I) = L_\mathrm{k}(0)\left(1 + \frac{I^2}{2I_\ast^2}+ \dots \right), 
\end{equation}
where $L_\mathrm{k}(0)$ is given by Eq.~\eqref{Lk at T=0}, and $I_\ast$ is of the order of the deparing current \cite{Zmuidzinas_2012, Annunziata_2010}. 
This estimation is valid at~$T~\ll~T_\mathrm{c}$, taking into account that the linear term must vanish due to symmetry considerations.

A nonlinearity of grAl opens up a possibility to use a superconducting transmission line made of grAl for parametric interaction of microwave signals. For this study we fabricated a coplanar transmission line 78~mm long with a central strip width of 3.5~$\mu$m and a gap of 5~$\mu$m, on an oxidized silicon substrate (see Fig.~\ref{fig:2}c). We used a film of thickness $t$~=~30~nm with normal state sheet resistance $R_\mathrm{n}^\square$ = 200~$\Omega$ and critical temperature $T_\mathrm{c}$~=~2.15 K. Permittivity of our substrates was $\epsilon_\mathrm{r}$~=~11.9. The input and output of the transmission line were connected to the contact pads via impedance converters. 
 A characteristic impedance of a coplanar is 
\begin{equation}
\label{eq:3}
Z = \sqrt{(L_\mathrm{kl}+L_\mathrm{gl} )⁄C_\mathrm{l}},
\end{equation}
where $L_\mathrm{kl}$ and $L_\mathrm{gl}$ are geometric and kinetic inductances per unit length and $C_\mathrm{l}$ is a capacitance per unit length.

A phase velocity in a transmission line can be written as
\begin{equation}
\label{eq:4}
V_\mathrm{ph} = \frac{1}{\sqrt{(L_\mathrm{kl}+L_\mathrm{gl} )C_\mathrm{l}}}.
\end{equation}

Using an experimental setup shown in Fig.~\ref{fig:2}a  transport properties of our samples were measured. The grAl transmission lines were placed in a dilution refrigerator Oxford Instruments Triton DR200 at temperature of 35~mK. A sample (Fig.~\ref{fig:2}c) was placed in a Copper sample holder (Fig.~\ref{fig:2}b) and bonded by Aluminum wires with a diameter of 25~$\mu$m. A sample holder was thermally anchored to a mixing chamber plate of a refrigerator. 
Microwave signals were generated by Vector Network Analyzer Agilent Technologies PNA-X N5242A. An input signal was attenuated for reduction of thermal noises. In order to measure a nonlinearity of grAl we used a Mini-Circuits Bias-Tee ZX85-12G-S+ for injection of a dc bias current from a current source Keithley 6221.

\section{Results and discussion} 
In order to demonstrate the nonlinear properties of superconducting films, the phase of the microwave signal is measured as a function of the amplitude of the bias current. In approximation of low currents the nonlinear phase shift is proportional to the total phase length of the transmission line \cite{Adamyan_2016}:

\begin{equation}
\label{eq:5}
\phi(I) = 2\pi f l/V_\mathrm{ph}(I).
\end{equation}
In the limit of a high kinetic inductance $L_\mathrm{gl}\ll L_\mathrm{kl}$, knowing a current dependence of a phase shift $\Delta\phi(I)$ we can find a current dependence of a kinetic inductance using Eqs.~\eqref{eq:4} and \eqref{eq:5}:  
\begin{equation}
\label{eq:6}
L_\mathrm{k}(I) = \left(\sqrt{L_\mathrm{k}(0)}+\frac{\Delta\phi(I)}{2\pi f l\sqrt{C_\mathrm{l}}}\right)^2,
\end{equation}
where $L_\mathrm{k}(0)\approx$~3.63~H/m  is a kinetic inductance per unit length at low current defined by Eq.~(\ref{Lk at T=0}), $l$ is a total physical length of a transmission line, $C_\mathrm{l} = 1.25\times 10^{-10}$~F/m is a capacitance per unit length.

\begin{figure}[H]
\centering
\includegraphics[width=0.48\textwidth]{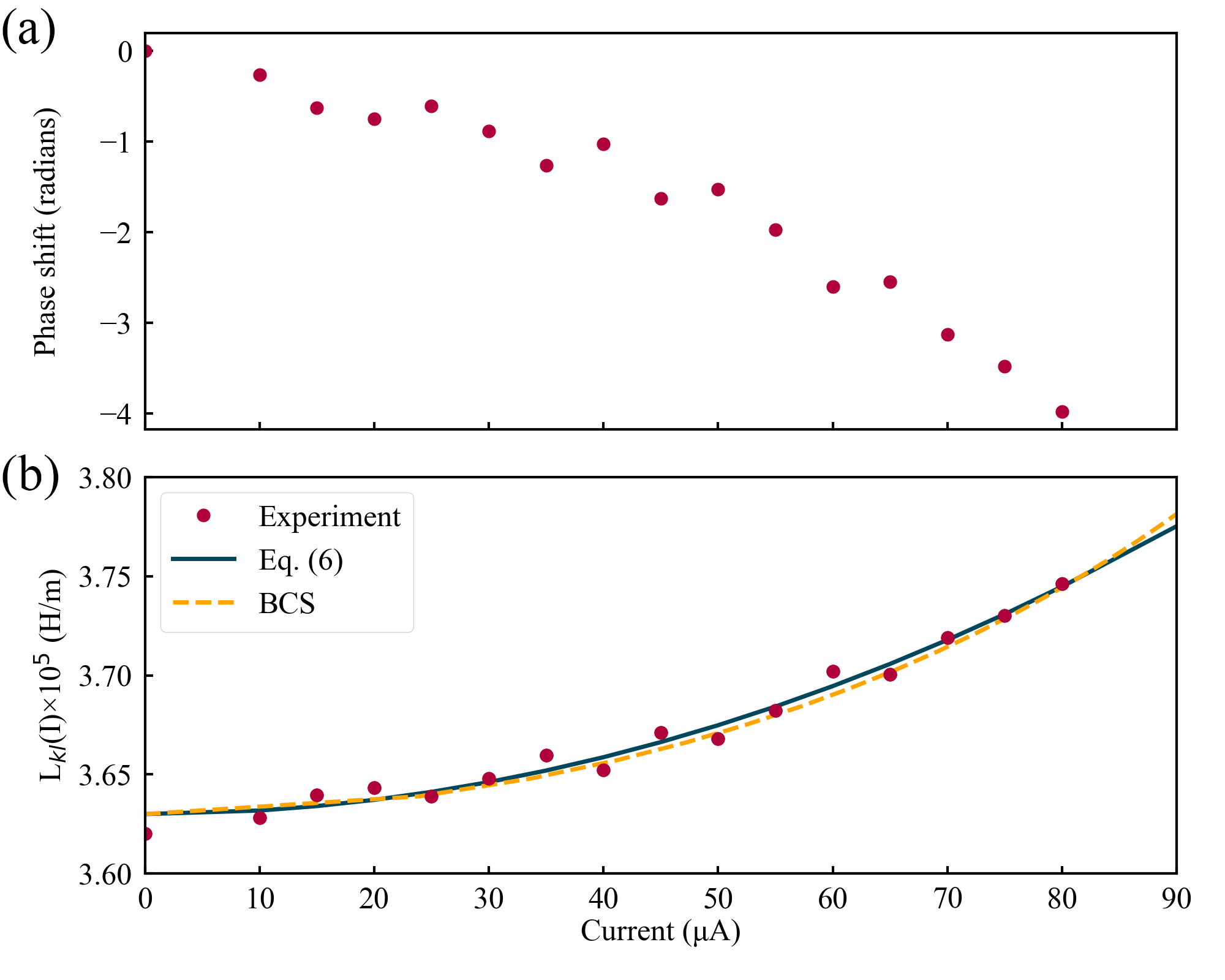}
\caption{\label{fig:3} The microwave signal's phase shift $\Delta\phi$ (a) and the kinetic inductance (b)  versus dc current  at a frequency of 7~GHz. At this frequency the largest phase shift is 4~radians, with total line phase length of 284~radians. Experimentally obtained kinetic inductance (red dots) is in a good agreement with both analytical curves: BCS approximation given by Eqs.~\eqref{Lk(I)} and \eqref{I/Ic} - a dashed orange line, and an approximation of Eq.~\eqref{eq:2} - a solid blue line.}
\end{figure}

Figure~\ref{fig:3} summarizes our results of the grAl nonlinear properties study. Red dots in Fig.~\ref{fig:3}a show the dependence of the microwave signal phase shift on the current at a frequency of 7~GHz. At the bias current of 80~$\mu$A the phase shift reaches its maximum $\Delta\phi = $ 4~radian with the total line phase length of 284~radians. This results in the nonlinearity $\Delta\phi/\phi$ of about 1.4~\%, which is comparable to values previously obtained on NbTiN films \cite{Zmuidzinas_2012, Adamyan_2016}. In Fig.~\ref{fig:3}b red dots show the kinetic inductance calculated from the measured phase shift using Eq.~\eqref{eq:6}.
 
An analytical dependence using an approximation~\eqref{eq:2} is shown in Fig.~\ref{fig:3}b by solid blue line, here $I_{\ast}~\approx$~318~$\mu$A. At the same figure by dashed orange line an analytical dependence from Eqs.~\eqref{Lk(I)} and \eqref{I/Ic} is shown, where $I_\mathrm{dep} \approx$~144~$\mu$A. 
Measurement results are in a good agreement with both analytical curves. A measured critical current $I_\mathrm{c}~\approx$~80~$\mu$A, which is 56~\% of the theoretical limit. We believe that the measured value of the critical current could be reduced by defects and not optimal structure geometry.
For a trivial long dispersionless line, a gain $G$ can be estimated as \cite{Zmuidzinas_2012}
\begin{equation}
\label{eq:7}
G \approx 1+(\Delta\phi)^2.
\end{equation}
In our experiments, the phase shift is caused by a dc bias current passing through the transmission line, which also occurs when passing a microwave signal of higher power. Using  Eq.~(\ref{eq:7}) we estimate a microwave signal amplification in a dispersionless transmission line to be expected of the order of 17~dB at a frequency of 7~GHz.

In conclusion, we experimentally demonstrated the nonlinear dependence of the grAl thin films kinetic inductance on the dc current, which, to the best of the authors' knowledge, has not been previously studied. On a line with a phase length of 284 radians, a phase shift of 4~radians is obtained at a frequency of 7~GHz. This value allows us to expect a significant amplification of the microwave signal, opening up the possibility of creating parametric amplifiers based on grAl.
\section*{Acknowledgements}
We thank V.V. Ryazanov, A.V. Karpov and A.V. Ustinov for supporting this research and giving valuable recommendations during the work on it.
The study of the granular Aluminum kinetic inductance nonlinearity was funded by Russian Science Foundation (project 21-72-30026, https://rscf.ru/en/project/21-72-30026/). The review of research on materials with high kinetic inductance and TWPA realisations was funded by  Strategic Academic Leadership Program "Priority-2030" (NUST MISIS grant K2-2022-029).


\end{document}